\documentclass[12pt, preprint]{aastex}
\newcommand\lam{\lambda}

\newcommand\sig{\sigma}
\newcommand\omg{\omega}
\newcommand\prm{\prime}
\begin{document}
\title{Merging Rates of the First Objects and the Formation of First Mini-Filaments in Models with Massive Neutrinos}
\author{Hyunmi Song and Jounghun Lee}
\affil{Astronomy Program, FPRD, Department of Physics and Astronomy,
Seoul National University, Seoul 151-747, Korea\\
\email{hmsong@astro.snu.ac.kr, jounghun@astro.snu.ac.kr}}

\begin{abstract}
We study the effect of massive neutrinos on the formation and evolution of the 
first filaments containing the first star-forming halos of mass $M\sim 10^{6}M_{\odot}$ 
at $z\sim 20$. With the help of the extended Press-Schechter formalism, we evaluate 
analytically the rates of merging of the first star-forming halos into zero-dimensional 
larger halos and one-dimensional first filaments. It is shown that as the neutrino mass 
fraction $f_{\nu}$ increases, the halo-to-filament merging rate increases while 
the halo-to-halo merging rate decreases sharply. For $f_{\nu}\le 0.04$, the halo-to-filament 
merging rate is negligibly low at all filament mass scales, while for $f_{\nu}\ge 0.07$ the 
halo-to-filament merging rate exceeds $0.1$ at the characteristic filament mass scale of 
$\sim 10^{9}M_{\odot}$. The distribution of the redshifts at which the first filaments 
ultimately collapse along their longest axes is derived and found to have a sharp maximum 
at $z\sim 8$.  We also investigate the formation and evolution of the second generation 
filaments which contain the first galaxies of mass $10^{9}M_{\odot}$ at $z=8$ as the parent of 
the first generation filaments. A similar trend is found: For $f_{\nu}\ge 0.07$ the rate of clustering 
of the first galaxies into the second-generation filaments exceeds $0.3$ at the characteristic mass 
scale of $\sim 10^{11}M_{\odot}$. The longest-axis collapse of these second-generation filaments are 
found to occur at $z\sim 3$. The implications of our results on the formation of massive high-$z$ 
galaxies and the early metal enrichment of the intergalactic media by supernova-driven  
outflows, and possibility of constraining the neutrino mass from the mass 
distribution of the high-$z$ central blackholes are discussed.
\end{abstract}


\section{INTRODUCTION}

In 1970, Zel'dovich suggested that the large-scale structure of the Universe such as sheets and filaments form 
through the anisotropic collapse of large-scale density fluctuations along the principal axes of the gravitational 
tidal fields \citep{zel70}. Although his original model assumed a hot dark matter cosmology, this picture can be well 
accommodated even by the currently popular $\Lambda$-dominated cold dark matter ($\Lambda$CDM) cosmology \citep{web96}. 
The dark matter halos that condense out first on the smallest scales merge anisotropically along the principal 
axes of the tidal fields into larger and larger structures. The inherent anisotropic nature of the gravitational 
merging process leads to the formation of filament-like and sheet-like structures, as envisioned by 
Zel'dovich's model. The nonlinear evolution of the tidal shear fields sharpens the anisotropic interconnection 
of the large-scale structure (which is often called the cosmic web), developing a hierarchical system where small 
filaments are threaded with the larger parent filaments \citep[e.g,][]{springel-etal05}. To understand the formation 
and evolution of the large-scale structure, an important question to answer is how frequently the halos merge into 
filaments and under what circumstances the halo-to-filament merging event occurs efficiently.

\citet{LC93,LC94} have for the first time evaluated analytically the merging rates of bound halos by extending the 
original Press-Schechter formalism \citep{PS74}. Their analytic work was later refined and complimented by 
numerical calculation of merger trees from N-body simulations \citep{somerville-etal00}.  Recently, \citet{FM08} 
constructed merger trees from the large Millennium Run halo catalogs \citep{millennium} and found that the mean 
merging rate per halo can be well modeled by a nearly universal fitting formula. In addition, \citet{FM08} showed 
that the prediction of the extended Press-Schechter formalism for the halo merging rates disagrees with the
numerical results from the Millennium Run halo catalog by up to a factor of a few. 

The halo-to halo merging rates were originally derived under the assumption that the merging events 
occur in a hierarchical way. If the matter content of the Universe were composed purely of CDM particles, then the 
structure formation would proceed in a strictly hierarchical way.  However,  if the non-CDM particles (either warm or 
hot) coexist with the CDM particles, the pure hierarchy in the build-up of the large-scale structure would 
break down at some level, depending on their mass fraction \citep{yoshida-etal03}. True as it is that 
the current observations put a tight constraint on the amount of the possible non-CDM particles, they
have not been completely excluded. As a matter of fact, the existence of non-CDM particles have been suggested as 
one possible solution to alleviate the apparent mismatches between observations and the predictions of 
$\Lambda$CDM cosmology on the subgalactic scales \citep[see][for a recent discussion]{boy-etal08,kuzio-etal10,DS10}.

Among the various candidates for the non-CDM particles, the massive neutrinos have so far attained the most serious 
attentions.  Ever since it was discovered that at least one flavor of the neutrinos must be massive and that the 
neutrino masses would be best constrained from the astronomical data \citep[see][for a review]{LP06}, 
a flurry of research has been conducted to quantify the effect of massive neutrinos on the formation and evolution 
of the large-scale structure 
\citep[e.g.,][]{bond-etal80,doroshkevich-etal81,hu98,HE98,valdarnini-etal98,EH99,LC02,seljak-etal06,
tegmark-etal06,saito-etal09,SK10}.  In previous studies of the structure formation in a $\Lambda$MDM 
($\Lambda$+CDM+massive neutrinos) universe,  what has been largely focused is the suppression of the 
small-scale powers in the density power spectrum due to massive neutrino's free-streaming and its effect 
on the structure formation. 

Since the massive neutrinos would decrease the merging rates of the small-scale halos due to their 
free streaming, one would naively think that the formation of large scale structures would be 
deferred in a $\Lambda$MDM universe. Meanwhile, the recent discovery of high-$z$ 
quasars at $z\ge 6$ \citep{fan-etal01,fan03,fan-etal04,fan-etal06,jiang-etal09,willott-etal09,willott-etal10} 
has suggested that the first galaxies of mass $\ge 10^{8}M_{\odot}$ 
must have formed at redshifts $z\ge 8$. At first glance the existence of such massive first 
galaxies at such early epochs might look difficult to reconcile with a $\Lambda$MDM model. 
Even in a $\Lambda$CDM cosmology, it is not so easy to explain the existence of the high-$z$ quasars  
since it requires very high merging/accretion rate in the early Universe. 

Yet, what has not been noted in previous works is that the presence of massive neutrinos may actually open 
a new channel for the rapid structure formation. Since the massive neutrinos would behave as effective warm 
dark matter (WDM) particles, they would fasten the collapse of matter along the major axes of the gravitational 
tidal fields on large scales. In other words, while the free-streaming of massive neutrinos slows down the 
formation of the zero-dimensional bound halos on small scales, it speeds up the formation of one-dimensional 
filaments on larger scales. The filaments are the most optimal environments for the rapid accretion of gas 
and matter. Furthermore, when the filaments collapse along their longest axes, they would provoke the 
violent merging of the constituent halos. 

Here, assuming a $\Lambda$MDM universe, we study how the first filaments form through the 
clustering of the first star-forming halos and how they evolve after the longest-axis collapse. 
The most idealistic approach for the study of the first filaments would be to use high-resolution 
N-body simulations for a $\Lambda$MDM universe. But, it is still a daunting task to implement the 
dynamics of the massive neutrinos into the current N-body simulations. Instead, we choose an analytic 
approach based on the extended Press-Schechter formalism \citep{PS74,LC93} for this study. 
Throughout this paper, we assume flat $\Lambda$MDM models where the neutrinos have mass and three 
species $N_{\nu}=3$. The key cosmological parameters are set at $\Omega_{m}=0.266,\ \Omega_{\Lambda}=0.734,\ 
\Omega_{b}=0.0449,\ \sigma_{8}=0.801$ (for the $\Lambda$CDM part), $h=0.710$  and $n_{s}=0.963$, 
to be consistent with WMAP7 results \citep{wmap7}. 

The outline of this Paper is as follows. In \S 2, we derive analytically the rates of merging of 
the first star forming halos into the first filaments and determine their characteristic mass scales.
In \S 3.1, we evaluate the epochs of the longest-axis collapse of the first filaments and the rates of 
merging of the first galaxies into the second-generation filaments, and determine their characteristic 
mass scales.  In \S 4, we explore the possibility of constraining the neutrino mass from the 
observable mass distribution of the high-$z$ supermassive blackholes. In \S 5, the results are summarized and 
a final conclusion is drawn.

\section{FORMATION OF THE FIRST FILAMENTS}

Recent observations indicate that the first galaxies must have formed in the halos 
of mass $10^{8}-10^{9}M_{\odot}$ at redshifts $z=9-10$ \citep[e.g.,][]{willott-etal09,willott-etal10}. 
What has been inferred from the current high-resolution hydrodynamic simulation is that these 
first galaxies may have formed through rapid merging of the first-star forming minihalos of mass 
$10^{6}M_{\odot}$ at $z\ge 20$ \citep[][and references therein]{bromm-etal09}. 
In this section, we investigate how the presence of massive neutrino would change the 
evolution channel of the first-star forming halos. 

\subsection{Variation of the Halo-to-Halo Merging Rates with the Neutrino Mass}

In the extended Press-Schechter theory developed by \citet{LC93,LC94}, the key 
quantity is the conditional probability, $f(M\prm,z\prm|M,z)$, that  
a halo of mass $M$ existing at a given redshift $z$ will merge to form a more 
massive halo of mass $M\prm (>M)$ at some lower redshift $z\prm(<z)$. 
Adopting the spherical collapse condition that a halo of mass $M$ forms at $z$ 
when its linear density contrast $\delta$ on the mass scale $M$ reaches a 
redshift-dependent threshold $\delta_{c}/D(z)$ where $\delta_{c}\approx 1.68$ 
and $D(z)$ is the linear growth factor, \citet{LC93} equated  this key 
conditional probability to the differential fraction of the volume in the 
linear density field occupied by the regions satisfying $\delta\prm \ge 
\delta_{c}/D(z\prm)$ on the mass scale $M\prm$ at $z\prm$ provided that it also 
satisfies the condition of $\delta=\delta_{c}/D(z)$ on the mass scale $M$ at 
earlier epoch $z$.  

This key conditional probability $f(M\prm,z\prm|M,z)\,dM\prm$ can now be written as 
\begin{eqnarray}
  f(S\prm,\omg\prm|S,\omg)\,dS\prm
  \label{eqn:start}
  &\equiv& 2\frac{d}{dS\prm} p(\delta_{M\prm}\ge\omg\prm|\delta_{M}=\omg)\,
dS\prm \nonumber \\
  &=& 2\frac{d}{dS\prm}\int_{\omg\prm}^{\infty}p(\delta_{M\prm}|\delta_{M}=\omg)
d\delta_{M\prm}\,dS\prm \nonumber \\
  &=&      \frac{1}{\sqrt{2\pi}}\left[\frac{S}{S\prm(S-S\prm)}\right]^{3/2}
           \frac{\omg\prm(\omg - \omg\prm)}{\omg}
\exp\left[-\frac{(\omg\prm S-\omg S\prm)^2}{2SS\prm(S-S\prm)}\right]\,dS\prm, 
  \label{eqn:con}
\end{eqnarray} 
where $f(M\prm,z\prm|M,z)\,dM\prm = f(S\prm,\omg\prm|S,\omg)\, dS\prm$ with
$S\equiv \sigma^{2}(M)$, $\omg\equiv \delta_{c}(z)$, 
$S\prm\equiv \sigma^{2}(M\prm)$, and $\omg\prm\equiv \delta_{c}(z\prm)$. 
Here $\sigma(M)$ and $\sigma(M\prm)$ are the rms fluctuation of the linear 
density field smoothed on the mass scale $M$ and $M\prm$, respectively.  
Normalization of $\sigma$ is done for the $\Lambda$MDM linear power spectrum 
to be identical with the $\Lambda$CDM linear power spectrum at large scale.
Here, of course, with the normalized $\Lambda$CDM linear power spectrum
$\sigma$ at $8\rm{h}^{-1}\rm{Mpc}$ has the value of $\sigma_8$ given by WMAP7.

Using Equation (\ref{eqn:con}), \citet{LC93} calculated the (instantaneous) rate of 
merging of a dark halo with mass $M$ at a given redshift $z$ into a larger halo with mass 
$M\prm$ by taking the derivative of $f(S\prm,\omg\prm|S,\omg)$ with respect to $z\prm$:
\begin{eqnarray}
  \frac{d^2 p(M\to M\prm|z)}{d{\rm ln}\Delta M dz}
  &=& -\left|\frac{dS\prm}{d{\rm ln}\Delta M}\right| 
\frac{d\omg\prm}{dz\prm}\frac{d}{d\omg\prm}\,
f(S\prm,\omg\prm|S,\omg)\,\bigg|_{z\prm=z} \\
  &=& -\frac{\Delta M}{{M\prm}^2} \,\bar{\rho}\, P_L (k\prm) 
\frac{d\omg}{dz}\frac{1}{\sqrt{2\pi}} 
\left[\frac{S}{S\prm (S -S\prm)}\right]^{3/2}\exp\left[-\frac{\omg^2}{2}
\left(\frac{1}{S\prm}-\frac{1}{S}\right)\right], 
\label{eqn:mrateH}
\end{eqnarray}
where $\Delta M \equiv M\prm-M$ and $P_{L}(k)$ is the linear density power spectrum for a 
$\Lambda$MDM cosmology. We use the analytic approximation given by \citet{HE98} for the 
evaluation of $P_{L}(k)$. 

With the help of the above analytic prescriptions, we determine the effect of 
massive neutrinos on the merging rate of the first star-forming mini-halos of 
mass $M=10^{6}\,M_{\odot}$ at redshift $z=20$ for the range of the neutrino mass 
fraction, $0.0\le f_{\nu}\le 0.13$.  We use the analytic formula given by \citet{HE98} 
for the $\Lambda$MDM linear density power spectrum 
$f_{\nu}\equiv \Omega_{\nu}/\Omega_{m}$ where $\Omega_{\nu}$ is the neutrino 
mass density parameter. The $\Lambda$MDM linear growth factor $D(z)$ 
is approximated as $D^{1-0.6f_\nu}_{\rm \Lambda CDM}(z)$ where $D_{\rm \Lambda CDM}(z)$ 
is the linear growth factor for a $\Lambda$CDM universe \citep{LP06}. 
The range of the neutrino mass fraction is chosen to be $0.0\le f_{\nu}\le 0.13$ 
where the upper limit of $0.13$ comes from the recent WMAP7 constraint \citep{wmap7}.
It is also worth mentioning here that for this range of $f_{\nu}$ the mini-halos of 
mass $10^{6}M_{\odot}$ at $z=20$ correspond to the $(3-4)\sigma$ peaks of 
the initial density field.

The rates of the merging of the first star-forming mini-halos are plotted as thin 
lines in Figure \ref{fig:merg1} for the five different cases of $f_{\nu}$, which 
reveals that the halo-to-halo merging rates decrease rapidly as $f_{\nu}$ increases.  
This result implies that in a $\Lambda$MDM universe the free streaming of the massive 
neutrinos have an effect of deferring the formation of larger halos through the 
major mergers of the first star-forming mini-halos.

\subsection{Variation of the Halo-to-Filament Merging Rate with the Neutrino Mass}

The result of \S 2 indicate that the direct merging of the first mini-halos into larger halos are slowed down by the free 
streaming of massive neutrinos. But, what about the merging of the first star-forming mini-halos into the first 
mini-filaments?  To derive the rate of merging of the first star-forming mini-halos into the mini-filaments, 
we adopt the classification scheme proposed by \citet{for-etal09}, according to whom a filament of mass 
$M\prm$ forms at $z\prm$ when the two of the three eigenvalues, 
$\lambda\prm_{1},\ \lambda\prm_{2},\ \lambda\prm_{3}$, of the linear deformation tensor 
(defined as the second derivative of the linear peculiar potential) smoothed on the scale of 
$M\prm$ exceed the threshold value, $\lambda_{c}/D(z\prm)$ with $\lambda_{c}\approx 0.3$. 
Using the N-body results, they demonstrated that the structures identified by this dynamical classification 
scheme exhibit indeed filamentary shapes. Their filament-finding scheme is also consistent with the 
statistical results obtained by \citet{LS98}. 

Adopting this condition for the formation of filaments, we modify the Lacey-Cole 
formalism to determine the halo-to-filament merging rates.
In this modified formalism, the key quantity is the conditional probability,  
$f(M\prm,z\prm|M,z)$, that a halo of mass $M$ observed at $z$ will 
merge into a filament of mass $M\prm$ formed at $z\prm$. We equate this 
conditional probability to the differential volume fraction occupied by the 
regions satisfying the filament formation condition, 
$\lambda\prm_1\ge\lambda\prm_2\ge\lambda_{c}/D(z\prm), 
\lambda\prm_3< \lambda_{c}/D(z\prm)$ on the mass scale $M\prm$ at redshift $z\prm$ 
provided that the linear density contrast of the same regions, $\delta$,  
on the smaller mass scale $M$ at earlier epoch $z$ satisfies the spherical 
collapse condition $\delta=\delta_{c}/D(z)$. Here $\delta$ equals the 
sum of the three eigenvalues of the deformation tensor smoothed 
on the mass scale $M$.

Now, this key conditional probability can be expressed as 
\begin{equation}
\label{eqn:ffh}
f(S\prm,\eta\prm|S,\omg)\,dS\prm = 
2\frac{d}{dS\prm}p(\lam\prm_1,\lam\prm_2\ge\eta\prm,\lam\prm_3<\eta\prm|
\delta=\omg)\,dS\prm,
\end{equation}
with $\eta\prm\equiv\lambda_{c}/D(z\prm)$. 
To evaluate $f(S\prm,\eta\prm|S,\omg)\,dS\prm$ in Equation (\ref{eqn:ffh}), 
it requires us the joint conditional probability distribution of 
$p(\lambda_{1},\ \lambda_{2},\ \lambda_{3}|\delta)$, which has been already 
derived by \citet{lee06} from the original work of \citet{doroshkevich70} as  
\begin{eqnarray}
p(\lam\prm_1,\lam\prm_2,\lam\prm_3|\delta)&=& 
\frac{p(\lam\prm_1,\lam\prm_2,\lam\prm_3,\delta)}{p(\delta)} \nonumber \\
  &=& \frac{3375}{8\sqrt{5}\pi}\frac{\sig}{\sig_\Delta {\sig\prm}^6}
      \,(\lam\prm_1-\lam\prm_2)(\lam\prm_2-\lam\prm_3)(\lam\prm_1-\lam\prm_3) 
\times \nonumber \\
  & & \exp\left[-\frac{1}{2\sig_\Delta^2} \left(\frac{\sig\prm}{\sig}\delta-
  \frac{\sig}{\sig\prm}I_1\right)^2\right]
      \exp\left[-\frac{5}{2{\sig\prm}^2}({I\prm_1}^2-3I\prm_2)\right],
\label{eqn:lee06}
\end{eqnarray}
where $\sig\equiv\sig(M)$,$\sig\prm\equiv\sig(M\prm)$, $\sig_\Delta\equiv\sqrt{\sig^2-{\sig\prm}^2}$, 
$I\prm_1\equiv\lam\prm_1+\lam\prm_2+\lam\prm_3$ and 
$I\prm_2\equiv\lam\prm_1\lam\prm_2+\lam\prm_2\lam\prm_3+\lam\prm_3\lam\prm_1$.

It is now straightforward to calculate the (instantaneous) rate of merging 
of the first star forming halos of mass $M=10^{6}\,M_{\odot}$ at $z=20$ 
into the early mini-filaments of mass $M\prm$ as  
\begin{eqnarray}
  \frac{d^2 p(M\to M\prm|z)}{d{\rm ln}\Delta M dz}
  &=& -\frac{\Delta M}{{M\prm}^2} \,\bar{\rho}\, P_L (k\prm)
       \frac{d}{dz\prm}\,f(S\prm,\eta\prm|S,\omg)\,\bigg|_{z\prm=z}.
\end{eqnarray}
Figure \ref{fig:merg1} plots the halo-to-filament merging rates (thick lines). In contrast 
to the halo-to-halo merging rates, the halo-to-filament merging rates increase with $f_{\nu}$. 
As $f_{\nu}$ changes from $0$ to $0.13$, the maximum halo-to-filament merging rates increase 
by a factor of $3$. We call the scale on which the halo-to-filament merging rate reaches the 
maximum value the characteristic filament mass scale and denote it as $M_{F}$. 
For $f_{\nu}\ge 0.07$, the halo-to-filament merging rate begins to exceed 
$0.1$ at the characteristic mass scale around $M_{F}=10^{9}M_{\odot}$. 
Whereas for $f_{\nu}\le 0.04$, the filament-to-halo merging rate is below $0.1$ 
at all filament mass scale. Note that for $f_{\nu}=0.07$ the mini-halos of mass 
$10^{6}\,M_{\odot}$ at $z=20$ tend to merge more rapidly into the mini-filaments of mass 
$10^{9}\,M_{\odot}$ rather than into the larger halos of mass $10^{7}\,M_{\odot}$.
Figure \ref{fig:mf_fnu} plots the characteristic mass $M_{F}$ of the first filaments 
versus the neutrino mass fraction $f_{\nu}$. As one can see, the value of $M_{\nu}$ 
decreases almost linearly with $f_{\nu}$.

\section{EVOLUTION OF THE FIRST FILAMENTS}

\subsection{Longest-Axis Collapse of the First Filaments}

In the previous section, it is shown that the first star-forming mini-halos would merger 
rapidly into the first filament of characteristic mass $\sim 10^{9}M_{\odot}$ in the presence 
of massive neutrinos. As the Universe evolves, these first filaments would ultimately 
collapse along their longest-axes that are in the direction of the minor principal axes 
of the local tidal field. To determine the distribution of the redshift $z_{f}$ when the 
first filaments collapse along their longest axes, we calculate the conditional probability 
density that a region satisfying the filament-formation condition 
($\lam_1\ge\eta,\lam_2=\eta,\lam_3<\eta$) at $z=20$ on the mass scale $M_{F}$ will 
meet the halo-formation condition ($\delta=\omega$) at the same mass scale 
\footnote{To prevent the mathematical divergence, the mass of a final halo is set at 
$0.99M_{F}$.} but at some lower redshift, $z_{f}<20$:
\begin{equation}
  p(\delta\prm=\omg\prm|\lam_1\ge\eta,\lam_2=\eta,\lam_3<\eta)
  = \frac{\int_{-\infty}^{\eta}d\lam_3 \int_{\eta}^{\infty}d\lam_1 \,\, 
      p(\delta\prm=\omg\prm,\lam_1,\lam_2=\eta,\lam_3)}
      {\int_{-\infty}^{\eta}d\lam_3 \int_{\eta}^{\infty}d\lam_1 \,\, 
p(\lam_1,\lam_2=\eta,\lam_3)}, 
\label{eqn:epoch}
\end{equation}
where $\delta\prm$ is the linear density contrast on the mass scale $M_{F}$ at $z_{f}$, 
$\lambda_{1},\lambda_{2},\lambda_{3}$ are the three eigenvalues of the deformation tensor 
smoothed on the mass scale of $M_{F}$ at $z=20$, $\omg\prm\equiv\delta_{c}/D(z_{f})$ and 
$\eta\equiv\lambda_{c}/D(z)$. Using the joint probability density distribution of 
$p(\delta\prm,\lambda_1,\lambda_2,\lambda_3)$ derived by \citet{doroshkevich70,lee06}, 
it is also straightforward to evaluate equation (\ref{eqn:epoch}). 

Figure \ref{fig:epoch1} plots this distribution $p(z_{f})$ of the epochs of the 
longest axis collapse of the first filaments for the five different cases of $f_{\nu}$. 
As can be seen, $p(z_{f})$ has a maximum value at $z_{f}=8$-$9$ for all five cases, 
which implies that the first filaments retain their structures for a long period from $z=20$ to 
$z=8$-$9$ without collapsing into zero-dimensional halos. As $f_{\nu}$ increases, however, the 
distribution $p(z_{f})$ becomes narrower and its peak position moves slightly to the lower value 
of $z_{f}$. That is, the massive neutrinos play a role of delaying the longest-axis collapse of 
the first filaments. 

\subsection{Formation and Evolution of the Second-Generation Filaments}

Now that the halo-to-filament merging rates are found to increase with the neutrino mass fraction, 
we would like to see how the presence of the massive neutrinos alters the merging rates of the 
first galaxies. It is naturally expected that some fraction of the first galaxies would also 
merge first in the filaments (called the second-generation filaments). A question is what the 
characteristic masses of the second generation filaments are and when they would undergo the 
longest-axis collapse. To answer this question, we recalculate everything through equations 
\ref{eqn:start}-\ref{eqn:epoch}, but for the first galactic halos of mass $10^{9}M_{\odot}$ at $z=8$. 

\begin{figure}
\includegraphics[scale=1]{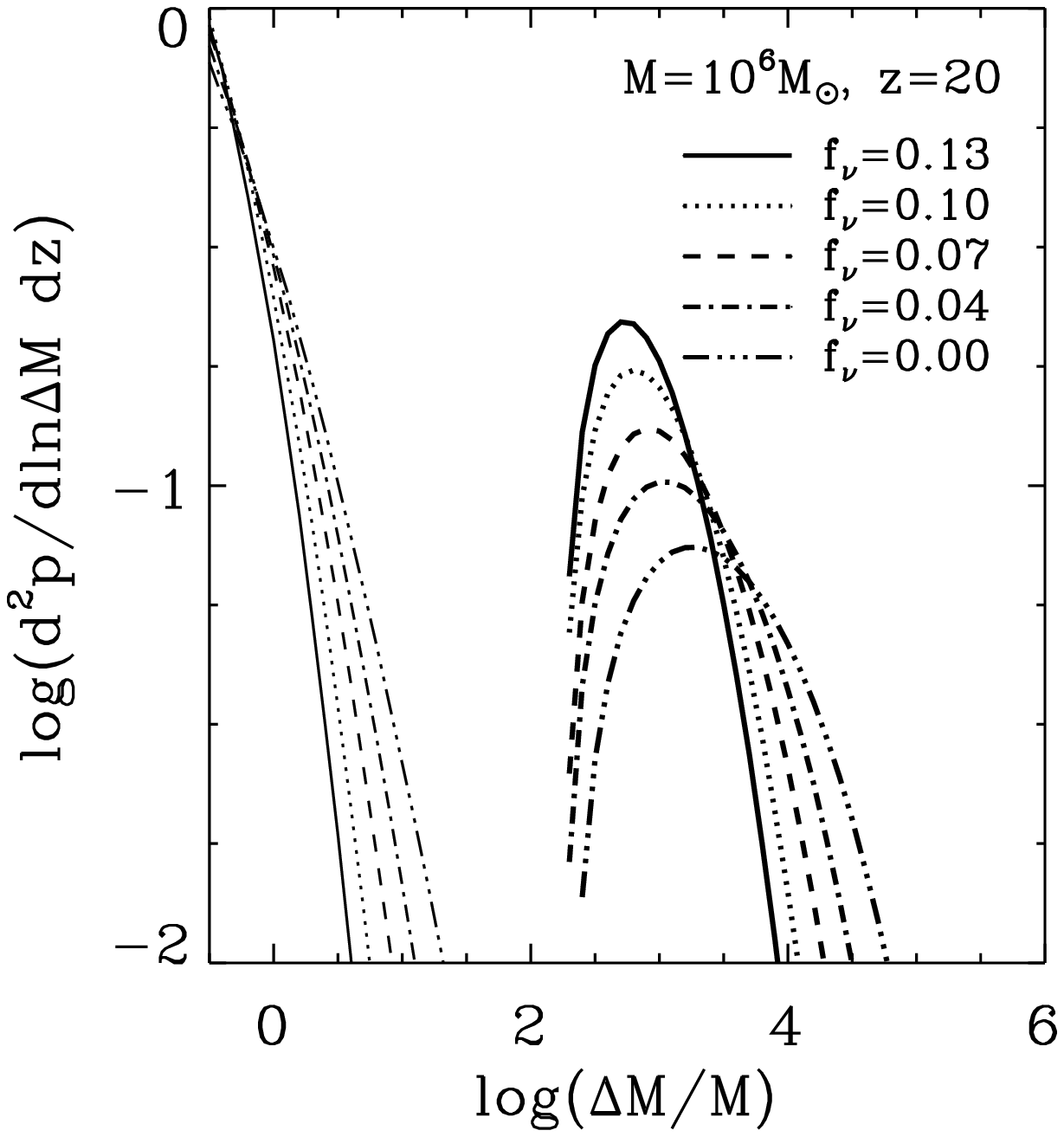}
\caption{Rates of the merging of the first star-forming halos with $M=10^{6}M_{\odot}$ 
into the first filaments (thick lines) and larger halos (thin lines) at $z=20$ 
for the five different values of the neutrino mass fraction: 
$f_{\nu}=0.13,\ 0.10,\ 0.07,\ 0.04$ and $0.0$ 
(solid, dotted, dashed, dot-dashed, and dot-dot-dashed, respectively).}
\label{fig:merg1}
\end{figure}

\begin{figure}
\includegraphics[scale=1]{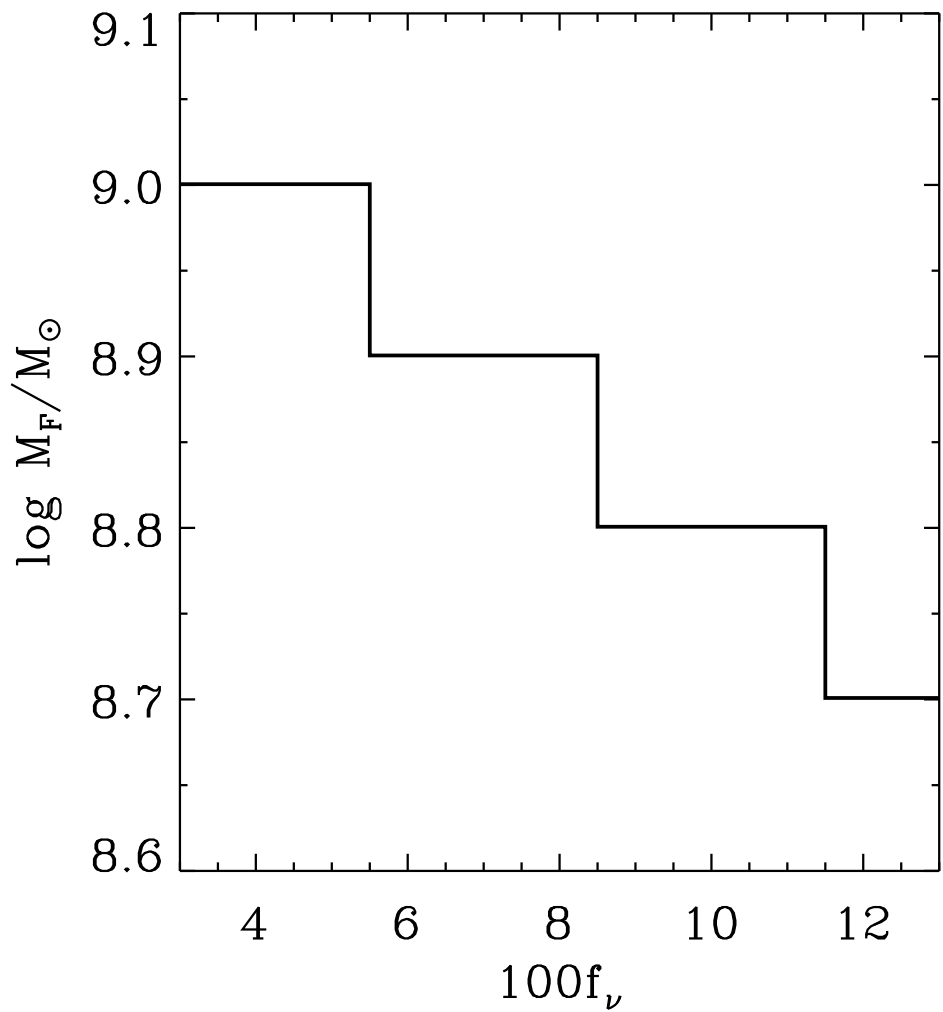}
\caption{Histogram for the characteristic masses of the first filaments 
versus the neutrino mass fraction $f_{\nu}$.}
\label{fig:mf_fnu}
\end{figure}

\begin{figure}
\includegraphics[scale=1]{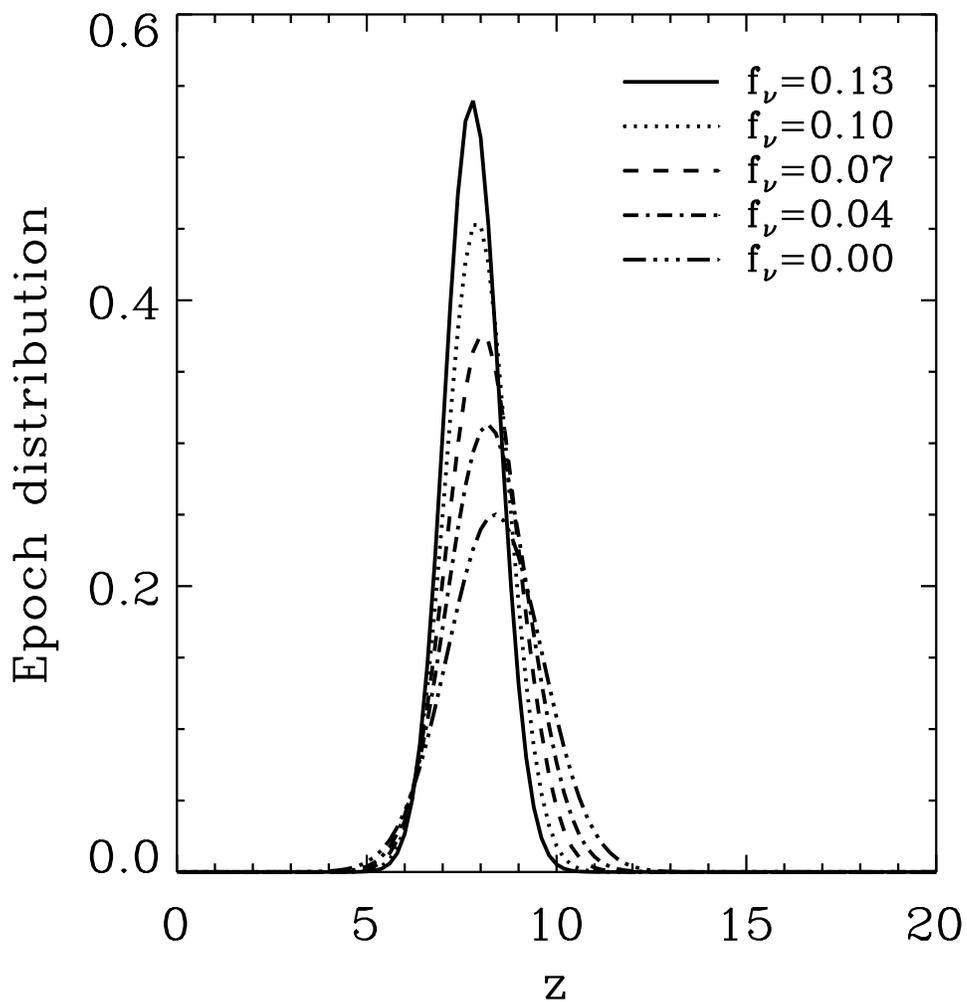}
\caption{Probability density distribution of the redshifts when the longest-axis collapse of the 
first filaments occur for the first different cases of the neutrino mass fraction: 
$f_{\nu}=0.13,\ 0.10,\ 0.07,\ 0.04$ and $0.0$ (solid, dotted, dashed, dot-dashed, 
and dot-dot-dashed, respectively).}
\label{fig:epoch1}
\end{figure}

\begin{figure}
\includegraphics[scale=1]{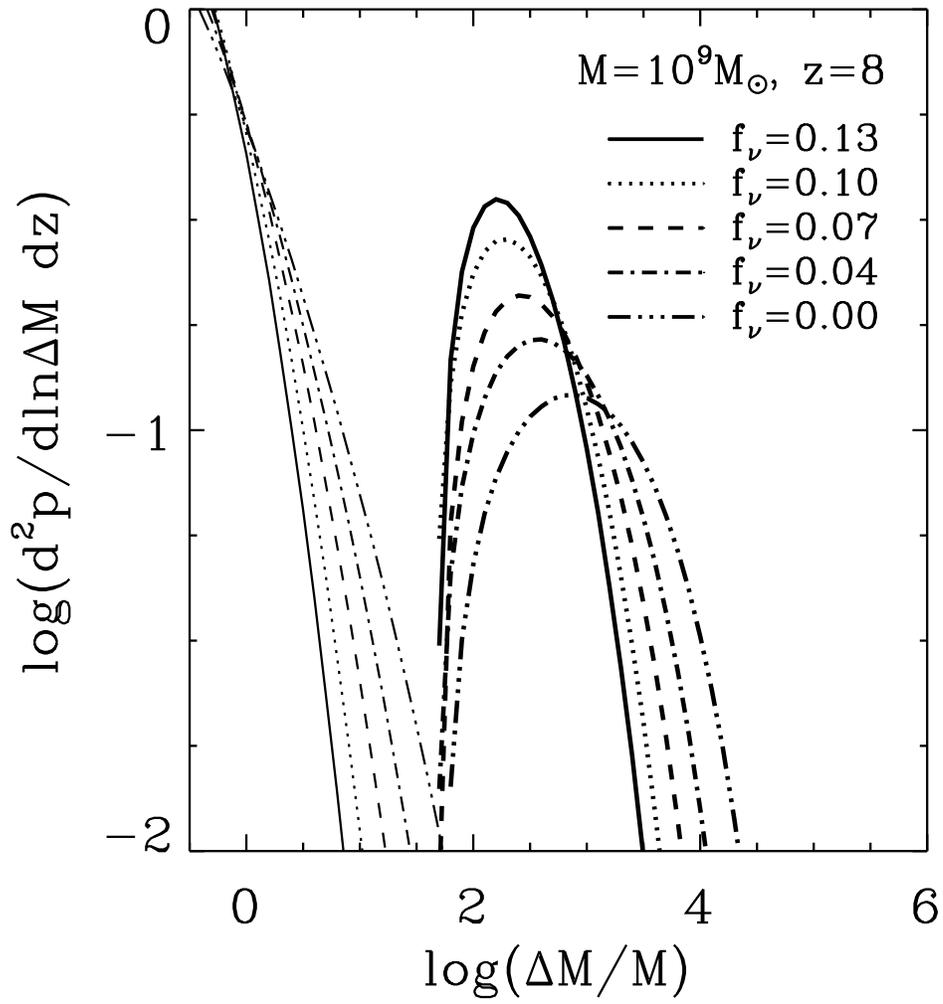}
\caption{Same as Figure \ref{fig:merg1} but for the first galactic halos of mass $M=10^{9}M_{\odot}$ 
at $z=8$.}
\label{fig:merg2}
\end{figure}

Figure \ref{fig:merg2} plots the halo-to-halo and halo-to-filament merging rates of the first galactic 
halos of mass $10^{9}\,M_{\odot}$ at $z=8$ as thin and thick lines, respectively. 
A similar trend to that shown in Figure \ref{fig:merg1} is found for the first galactic halos: 
As the neutrino mass fraction increases, the halo-to-halo merging rate decreases while the halo-to-filament 
merging rate increases. The comparison of Figs.\ref{fig:merg1} and \ref{fig:merg2} also reveals that the 
halo-to-filament merging rates increases with redshift and with halo mass. For $f_{\nu}\ge 0.04$, 
the rate of merging of the first galaxies into the second generation filaments exceeds $0.1$ at all 
mass scales. The characteristic mass of the second generation filaments is found to lie in range 
of $10^{11}<M/M_{\odot}\le 10^{12}$.  

\begin{figure}
\includegraphics[scale=1]{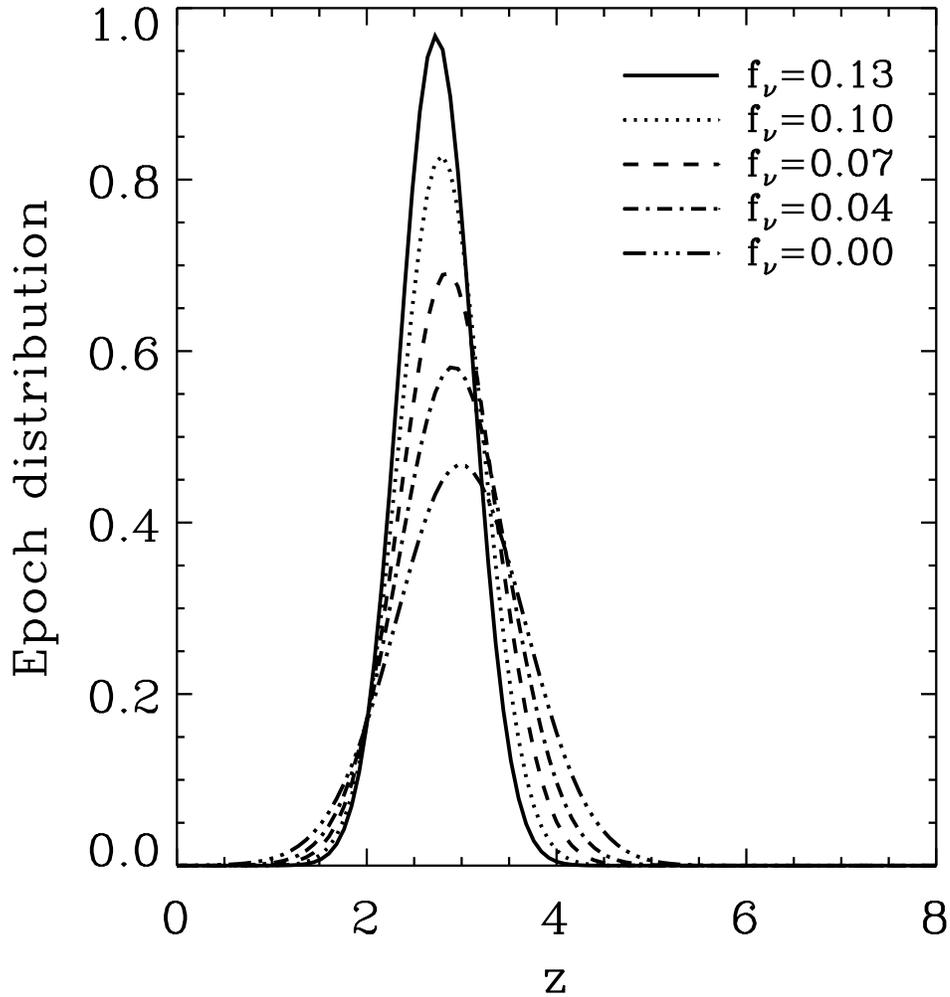}
\caption{Same as Figure \ref{fig:epoch1} but for the second-generation filaments of mass 
$M=10^{11}M_{\odot}$.}
\label{fig:epoch2}
\end{figure}

Figure \ref{fig:epoch2} plots the distribution of the redshifts at which the second generation 
filaments of mass $10^{11}M_{\odot}$ formed at $z=8$ experience the longest axis collapse. 
A similar trend to that shown in Figure \ref{fig:epoch1} is also found: $p(z_{f})$ has a maximum value 
at $z_{f}=3$-$4$ for all five cases, which implies that the second-generation filaments last 
from $z=8$ to $z=3$-$4$ without collapsing into zero-dimensional halos. The increase of $f_{\nu}$ 
results in narrowing down the shape of $p(z_{f})$ and shifting it slightly to the low-$z$ 
section. Note that these second generation filaments correspond to filaments hosting galactic halos and
are expected to be interconnected to the larger-scale parent filaments of mass $\ge 10^{13}-10^{14}M_{\odot}$ 
\citep[e.g.,][]{springel-etal05}. The massive galaxies formed through the longest-axis collapse of the 
second generation filaments would grow through the anisotropic accretion of matter and gas along 
the larger-scale parent filaments in the cosmic web. 

\section{CONSTRAINING THE NEUTRINO MASS WITH THE FILAMENT ABUNDANCE}

Once the first filaments form through the merging of the first star-forming halos at $z\le 20$, the first stars 
comprising the first filaments would grow more rapidly as the accretion of matter and gas occur more efficiently 
along the bridges of the first filaments \citep{PL09}. When the first filaments collapse along their longest axes 
at $z\sim 8$, the strong collisions between stars and gas particles in the filaments would result in 
seeding the supermassive blackholes in the massive first galaxies that are believed to power the 
ultra-luminous high-$z$ quasars \citep{willott-etal10}.  In this scenario, the high-end slope of the mass distribution 
of the supermassive blackholes inferred from the observable luminosity function of the ultra-luminous high-$z$ 
quasars would reflect the mass distribution of the first filaments. Since the mass distribution of the first 
filaments depend sensitively on the mass of massive neutrinos, it might be possible to constrain the mass of 
neutrinos by comparing the mass distributions of the first filaments with that of the high-$z$ supermassive 
blackholes. 

The mass distribution function of the first filaments ($dN_{\rm F}/dM$) at the epoch of longest-axis collapse
($z_{f}$) can be derived by employing the Press-Schechter approach \citep{PS74}:
\begin{equation}
\frac{dN_{\rm F}(M,z_{f})}{dM} = \frac{\bar{\rho}}{M}\left\vert\frac{d}{dM}F(M,z_{f})\right\vert, 
\label{eqn:mf}
\end{equation} 
where $\bar{\rho}$ is the mean mass density. Here, $F(M,z_{f})$ represents the volume fraction occupied by the first 
filaments of mass larger than $M$ at $z_{f}$. Following the Press-Schechter approach, this volume fraction can be 
expressed as 
\begin{equation}
\label{eqn:fm}
F(M,z_{f}) = A\ p[\lambda_{1}\ge\lambda_{2}\ge \lambda_{c,z_f},\lambda_{3}\le \lambda_{c,z_f} |\sigma_{M}],
\end{equation}
where $A$ is the normalization factor, $\lambda_{s}$ is the value of the $\lambda_{2}$ at the epoch 
of the longest-axis collapse.  Since the  first filaments formed at the threshold of $\lambda_{c}=0.3$,  
the value of $\lambda_{2}$ increases gradually till they collapse eventually along their longest axes. 
It has been found that at the epoch of the longest-axis collapse, $\lambda_{2}$ reaches the critical value of 
$\lambda_{s}=1.0$. 

\begin{figure}
\includegraphics[scale=1]{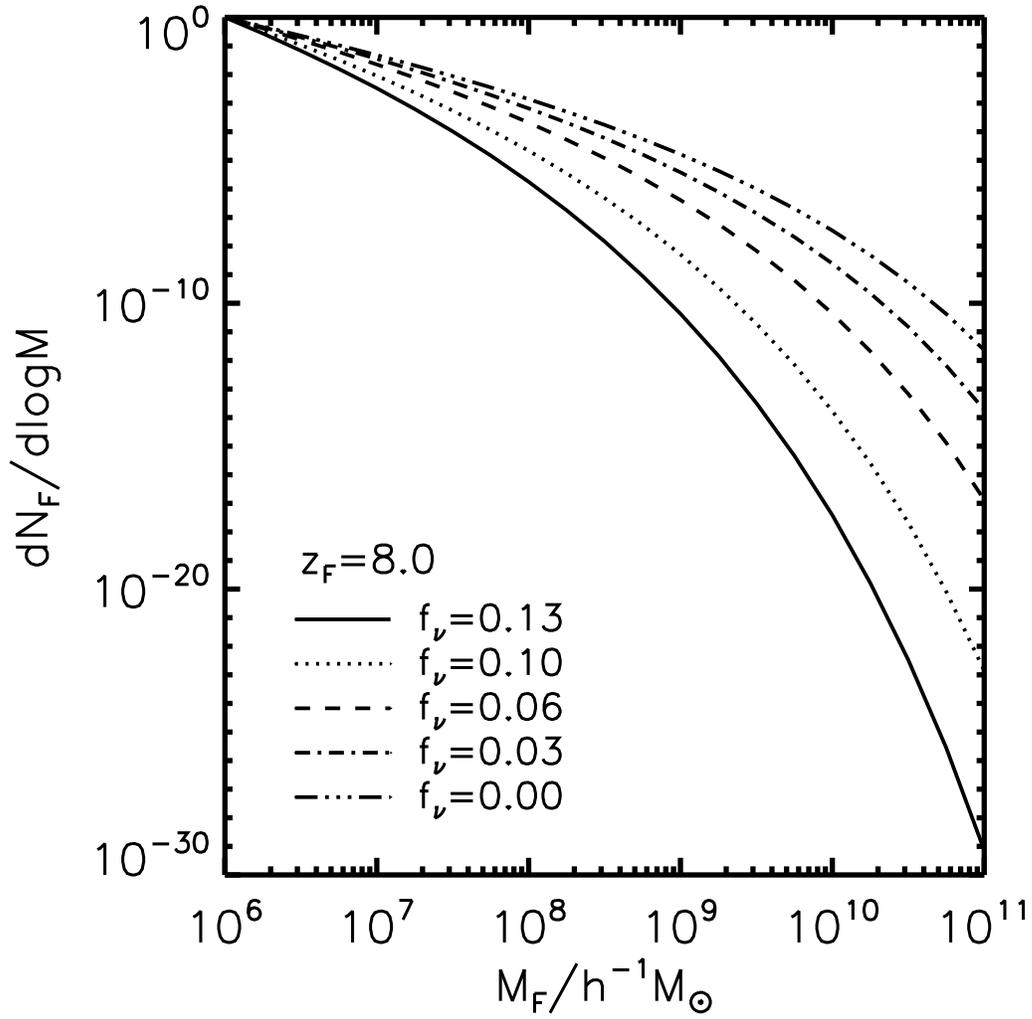}
\caption{Mass distribution of the first filaments for five different cases of the neutrino mass fraction $f_{\nu}$.}
\label{fig:mf}
\end{figure}

Using the joint distribution, of $p(\lambda_{1},\ \lambda_{2},\ \lambda_{3})$, derived by \citet{doroshkevich70}, 
it is straightforward to evaluate Equation (\ref{eqn:mf}). Figure \ref{fig:mf} plots the mass distribution 
of the first filaments in logarithmic scales at the epoch of the longes-axis collapse for five different cases of the 
neutrino mass fraction, $f_{\nu}$.  As we have already shown in \S 2, that the longest-axis collapse of the first 
filaments occur around $z=8$, we set $z_{f}$ at $8$ for the evaluation of Equation (\ref{eqn:mf}). 
As it can be seen, in the high-mass section ($M\ge 10^{9}\,h^{-1}M_{\odot}$), the abundance of 
the first filaments decreases sharply with $M$ for all cases. The rate of the decrease of the mass function of 
the first filaments, however, depends sensitively on the value of $f_{\nu}$.  The larger the value of $f_{\nu}$ is, 
the more rapidly the mass function drops in the high-mass section.  

\begin{figure}
\includegraphics[scale=1]{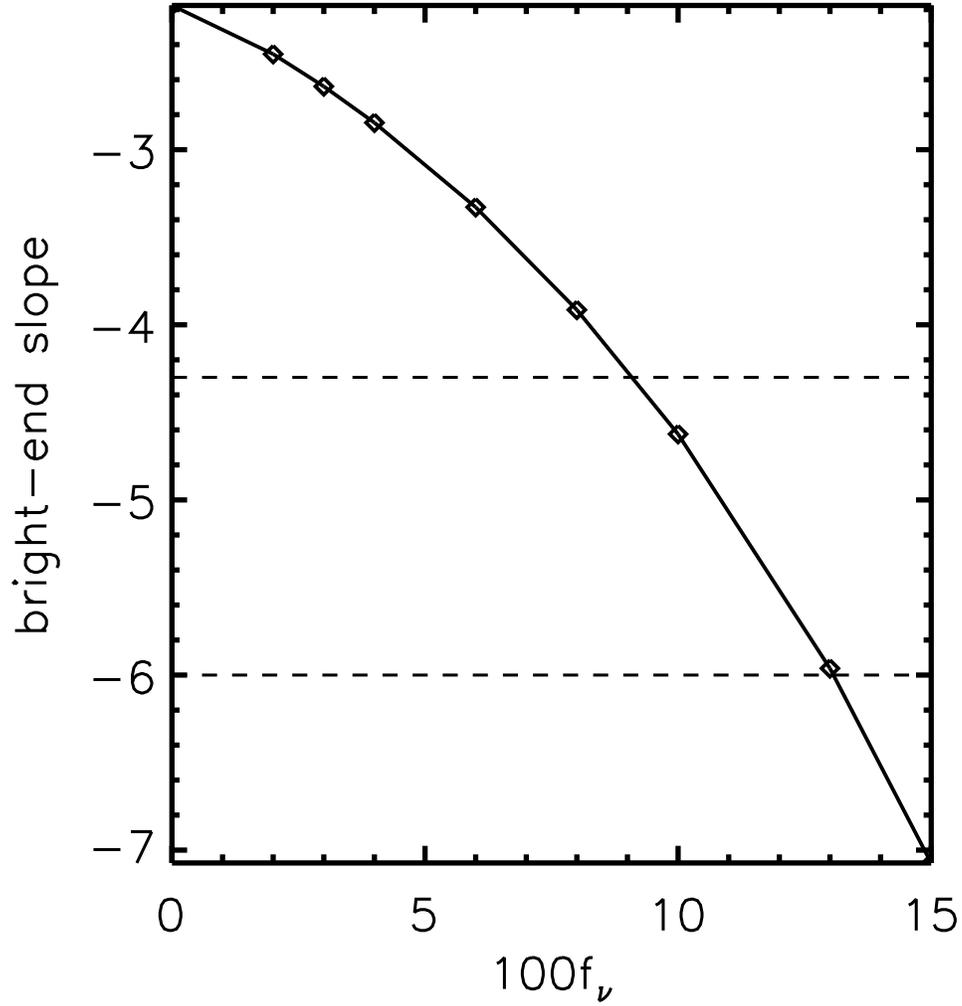}
\caption{High-end slope of the mass function of the first filaments versus $f_{\nu}$ at $z_{f}=8$ (solid line).  
Observational constraints from the mass function of the high-$z$ supermassive blackholes from 
\citet{willott-etal10} are shown as dashed lines.}
\label{fig:slope}
\end{figure}

To examine quantitatively how the rate of the decrease of the mass function of the first filaments in the 
high-mass section changes with $f_{\nu}$, we approximate it as a power law $M^{\alpha}$ and determine the value 
of the high-end slope, $\alpha$, in the mass range of $10^{9}\le M/(h^{-1}M_{\odot})\le 10^{11}$.
Figure \ref{fig:slope} plots the high-end slope, $\alpha$, of the mass distribution of the first filaments versus 
the neutrino mass fraction, $f_{\nu}$ as solid line.  As it can be seen, the absolute value of the high-end slope of 
the mass function of the first filaments increases sharply as $f_{\nu}$ increases. This result suggests that the 
value of $f_{\nu}$ could be constrained by using the high-end slope of the mass function of the 
first filaments. In practice, of course, it is not possible to measure directly the high-end slope of the mass function 
of the first filaments. As mentioned above, however, we can infer it from the observable high-$z$ blackhole 
mass function, under the assumption that the longest-axis collapse of the first filaments with mass larger than the 
characteristic mass ($10^{9}\,h^{-1}M_{\odot}$) would result in the first galaxies hosting the supermassive 
blackholes that power the ultra-luminous high-$z$ quasars. 

Now, we would like to compare the high-end slope of this theoretically derived mass distribution of the first filaments 
with that of the observationally obtained mass function of the massive galaxies hosting the high-$z$ supermassive 
blackholes. Very recently, \citet{willott-etal10} have derived the blackhole mass function at $z\sim 6$ from the 
observed luminosity function of the ultra-luminous high-$z$ quasars (see Figure 8 in Willott et al. 2010).   
Converting the mass of the high-$z$ supermassive blackholes into the total mass of their host galaxies through 
the following relation given by \citet{bandara-etal09}, 
$\rm{log}(M_{bh}/M_{\odot})=(8.18\pm0.11)+(1.55\pm0.31)\times(\rm{log}(M_{tot}/M_{\odot})-13.0)$, 
we find the best-fit high-end slope  is in the range of $-6\le \alpha\le -4.3$, which are plotted as two dashed lines 
in Figure \ref{fig:slope}. Comparing this range of $\alpha$ against the theoretical curve plotted in 
Figure \ref{fig:slope}, it is found that the neutrino mass fraction is constrained as $0.09\le f_{\nu}\le 0.13$ 
which corresponds to the neutrino mass range of $0.375\le m_{\nu}/(eV)\le 0.542$. 

\section{DISCUSSION AND CONCLUSIONS}

Starting with the first star-forming halos of mass $10^{6}M_{\odot}$ at $z=20$, we have shown analytically 
that the presence of massive neutrinos opens a new channel for the rapid formation of massive structures 
by speeding up the halo-to-filament merging process.  Once the first filaments of mass $\le 10^{9}M_{\odot}$ 
form through the clustering of some of the first star-forming halos at $z\ge 9$, matter and gas would accrete 
more efficiently along the first filaments, resulting in the fast growth of the constituent halos in the first 
filaments.  As the Universe evolves and the tidal forces increases, the first filaments would undergo the ultimate 
collapse into the zero-dimensional halos (corresponding to the first massive galaxies) in the directions 
parallel to the minor principal axes of the local tidal field.  

What has been found here is that the longest-axis collapses of the first filaments end up forming the first 
massive galaxies of mass $\le 10^{9}M_{\odot}$ at $z\ge 8$.  The longest axis collapses of the first 
filaments would accompany violent mergers among the constituent halos, which in turn would trigger 
the vigorous star-formation and feed the central blackholes in the resulting first galaxies. Our result suggests 
that the ultraluminous high-$z$ quasars detected at $z\ge 6$ might correspond to the first massive galaxies 
formed through the longest-axis collapses of the first filaments.

It has been also found that the second-generation filaments of mass $10^{9}M_{\odot}$ form through the 
clustering of some of the first galaxies of mass $10^{9}M_{\odot}$ at $z\sim 8$. The first galaxies in the second 
generation filaments would evolve fast through the filamentary accretion of matter and gas \citep{PL09}. 
When the second generation filaments collapse along their longest axes, they would also induce violent mergers 
among the constituent galaxies, which would lead to the formation of massive galaxies of mass larger than 
$10^{11}M_{\odot}$ with high star-formation rates and large central blackholes at redshifts as high as $z=3$. 

We test the possibility of constraining the neutrino mass, $m_{\nu}$, from the mass function of the high-$z$ 
supermassive blackholes that power the ultraluminous high-$z$ quasars, under the assumption that the longest 
axis collapses of the first filaments would seed the supermassive blackholes. 
Comparing the high-end slope of the theoretically derived mass function of the first filaments at the epoch of their 
longest-axis collapse with that of the observed mass function of the massive first galaxies inferred from the 
luminosity function of the ultra-luminous high-$z$ quasars, we find that the neutrino mass is constrained to be 
$0.375\le (m_{\nu}/eV)\le 0.542$.  

Yet, it is still only a feasibility study. A more comprehensive work should be done to put a robust constraint 
on $m_{\nu}$ from observational data. First of all, more refined theoretical models for the halo-to-filament merging 
rates and the abundance of the first filaments should be constructed than the crude approximations based on the extended 
Press-Schechter formalism, since the formalism has been shown to fail in accurately predicting the 
merging rates and abundance of bound halos \citep{FM08}. 
Second, the observational results from the luminosity function of the ultra-luminous high-$z$ quasars still suffer 
from small-number statistics and thus more quasar data are required. 
Third,  we use the relation given by \citet{bandara-etal09} to convert the blackhole mass into the halo mass. 
However, the relation of \citet{bandara-etal09} has been obtained from the blackholes detected at $z\le 4$, 
and thus it may not be applicable to the higher-$z$ blackholes. A more accurate relation between the 
high-$z$ blackhole mass and host halo mass should be found. 

Our results also hints that the massive neutrinos may have something to do with the early metal enrichment of the 
intergalactic medium (IGM). \citet{madau-etal01} showed that the early outflows driven by supernovae 
(SN) ejecta from the galaxies of mass $\ge 10^{8}M_{\odot}$ at $z\sim 9$ would enrich the IGM with the products of 
stellar nucleosynthesis provided that those galaxies possess relatively high star-formation efficiencies. 
It is interesting to note that the halo mass and epoch required for the SN-driven outflows are more or less coincident 
with our estimates for the galactic halos formed through the longest-axis collapse of the first filaments 
in a $\Lambda$MDM model. Assuming that the longest-axis collapse of the first filaments trigger star formation rapidly 
in the first massive galaxies in a $\Lambda$MDM model, the star-formation efficiency of the first mass galaxies may be 
as high as required for the early metal enrichment by SN-driven outflows. Furthermore, given our speculation 
that the longest-axis collapse of the first filaments would feed the supermassive blackholes, the anisotropic outflows 
from the active galactic nuclei (AGN) powered by the supermassive blackholes of the first massive galaxies might also 
contribute to the early metal enrichment of the IGM \citep[e.g.,][]{germain-etal09,barai-etal11}. 
It will be intriguing to explore how the metal enrichment of the IGM depends on the neutrino mass. Our future work is in 
this direction.

\acknowledgments

We thank an anonymous referee for helpful comments. 
This work was supported by the National Research Foundation of Korea (NRF) grant funded by the Korea 
government (MEST, No.2010-0007819). Support for this work was also provided by the National Research 
Foundation of Korea to the Center for Galaxy Evolution Research.

\newpage

\newpage

\newpage

\newpage

\newpage

\newpage

\newpage

\end{document}